**The Dynamics of a Highly Curved Membrane Revealed by All-atom Molecular Dynamics Simulation of a Full-scale Vesicle**

Christopher Kang,[+] Kazuumi Fujioka,[+] and Rui Sun[*]

fDepartment of Chemistry

The University of Hawai'i, Mānoa

2545 McCarthy Mall, Honolulu, HI 96817

Email: ruisun@hawaii.edu

Telephone: (808) 956-3207

[+]Authors contributed equally to the manuscript
[*]Author to whom correspondence should be addressed: ruisun@hawaii.edu

# Abstract

In spite of the great success that all-atom molecular dynamics simulations have seen in revealing the nature of the lipid bilayer, the interplay between a membrane's curvature and dynamics remains elusive. This is largely due to the computational challenges involved in simulating a highly curved membrane, as the one found in a small vesicle. In the present work, thanks to the computing power of Anton2, we present the first all-atom molecular dynamics simulation of a full-scale, realistically composed (both heterogeneous and asymmetric) vesicle of a meaningful time scale (over 10 microseconds), which reveals unique biophysical properties of various lipid molecules (diffusion coefficients, surface areas per lipid, order parameters) and packing defects in a highly curved environment. Most interestingly, a bilayer of the same lipid composition demonstrating no phase coexistence when flat shows very strong indictors of phase coexistence when highly curved. Lipid molecules found in the curvature-induced different phases are carefully verified by their distinct composition, area per lipid, parking defects, as well as diffusion coefficient. The result of the all-atom molecular dynamics simulations is consistent with previous experimental and theoretical models and enhance the understanding of nanoscale dynamics and membrane organization of small, highly curved organelles.

# I. Introduction

Membranes are integral components of every cell, providing the cell's identity and separating cellular compartments as well as their interior from the external medium. Besides acting as a physical barrier, membranes are now understood to be highly dynamic assemblies, changing their shape and composition based on what they compartmentalize. Biological membranes are intrinsically curved and their morphology has emerged as an integral regulator of cellular functions, including cell signaling[1], membrane fusion[2], and protein sorting.[3,4] Highly curved membranes are ubiquitous in intercellular vesicles,[5–7] intracellular structures such as mitochondria[8], Golgi[9], lipid droplets (LDs)[10,11] and the endoplasmic reticulum,[12] as well as protrusions such as microvilli[13], filipodia and invaginations (e.g., caveolae[1]). From the morphological point of view, membrane curvature imposes constraints on its constituents, therefore influencing local lipid composition that leads to various membrane dynamics.

Curvature has been known to induce certain lateral inhomogeneities in membranes.[14,15] One of the first examples is nanodomain phase separation, observed with freeze fracture electron microscopy over 40 years ago.[16] In recent years, there has been a growing interest in understanding the impact of membrane curvature on various cellular functions, one of which is lipid nanodomain organization (i.e., membrane tension, composition, curvature). Sorre and co-workers used a combination of fluorescence and force measurements to quantitatively show that a difference in lipid composition can build up between a curved and a non-curved membrane.[17] Leibler and Andelman[18] proposed a continuum model by which the acquisition of "spontaneous curvature", bending, and lipid composition provide a mechanism explaining the stabilization of phase-separated domains in heterogeneous membranes. Lipowsky and coworkers established an elastic membrane model to delineate the budding or invagination process[19] and emphasized the important role played by the curvature in relation to nanodomain stability.[20,21] This finding was later confirmed by experiments[22,23] as well as coarse-grained (CG) molecular dynamics (MD) simulations.[24–26]

Thanks to the revolution in computing power[27,28] and the development of force fields, all-atom (AA) molecular dynamics (MD) simulations have seen great success in understanding the nature of the cellular membrane – there are well established protocols to construct, validate, and simulate a *flat* membrane. Under the framework of periodic boundary conditions, the flat membrane in AA MD simulations expand to infinity in the lateral directions and have shown good agreement with experiments in area per lipid, thickness, etc.[29] Although a flat membrane is a reasonable approximation for studies where the membrane curvature is negligible (e.g., passive permeation, ion channels, etc.), it falls short in modeling biophysical processes that involve highly curved membranes. And on that front, unfortunately, there has been very little progress made in AA MD simulations over the past decade. For example, a *well-equilibrated* AA MD simulations of a highly curved membrane does not exist to demonstrate that the current force field parameters are able to model such system at all.

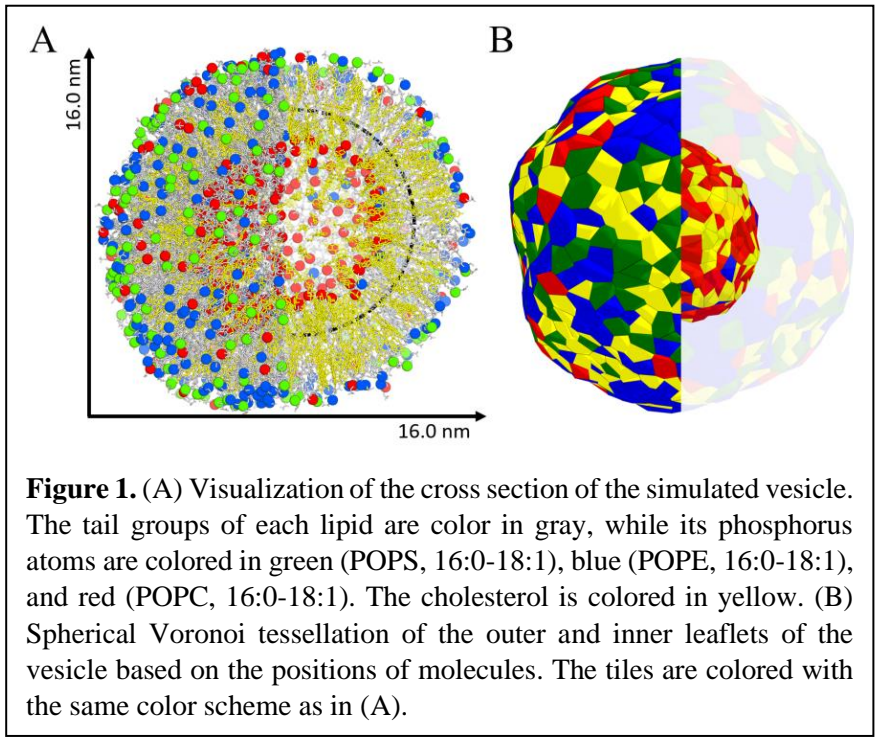

**Figure 1.** (A) Visualization of the cross section of the simulated vesicle. The tail groups of each lipid are color in gray, while its phosphorus atoms are colored in green (POPS, 16:0-18:1), blue (POPE, 16:0-18:1), and red (POPC, 16:0-18:1). The cholesterol is colored in yellow. (B) Spherical Voronoi tessellation of the outer and inner leaflets of the vesicle based on the positions of molecules. The tiles are colored with the same color scheme as in (A).

Such drought of study on curved membranes are primarily due to the efficiency of the simulations. Even for a vesicle of a diameter of ~20 nm (e.g., the size of the most minimal unit of membrane traffic[30]), such system would consist of over one million atoms including the solvating water. Unlike a flat membrane, simulating a patch of a vesicle will result in kinks at the boundary of the systems due to periodic boundary conditions[31]. To make the situation worse, previous AA MD situations have shown that it takes microseconds for the lipids to demonstrate proper phase behavior.[32] The size (e.g., ~ 1 million atoms) and temporal scales (i.e., microseconds) of the system quickly make the simulations computationally infeasable.[24] This issue could be fixed if one decreases the resolution of the simulation, e.g., employing CG[26] or mesoscopic models[33]. These models excel in recreating bulk-phase properties, but they lack information at the atomistic level and can sometimes lead to unphysical results.[10] To remain in the AA regime, considerable efforts have been made in mimicking membrane curvature, which in general can be classified into two categories: 1.) artificially maintaining membrane curvature using a bias force applied to curvature-related collective variables[34–36], dummy particles[37] or virtual walls[38] and 2.) altering membrane constituents, for example, introducing scaffolding proteins[39–41], packing different numbers of lipids on each leaflet[42], or introducing conical lipids[41,43–45]. Unfortunately, both categories are still yet to be substantiated by AA MD simulations and experiments. It is also important to note that these aforementioned methods can only study membrane surfaces with zero gaussian curvature (e.g., a microtubule) instead of a full sphere.

The present study reports on the first AA MD simulation of a realistically composed, full-scale, spherical synaptic vesicle (SV) that features both the heterogeneity and asymmetry of a bilayer at meaningful time scales (more than 10 μs). An SV is employed as a prototype because it is one of the smallest vesicles in the human body. The simulation sheds light on the dynamics of a highly curved membranes in the following aspects: (1) biophysical properties of various lipid molecules (diffusion coefficients, surface areas per lipid, order parameter, etc.), (2) the packing defects, and (3) the phase separation. Although being a simplified model of an actual SV in human neurons, the complete spherical structure is the first of its kind studied by AA MD and mimics highly curved membrane found in a small vesicle, thus providing unprecedented information at an atomistic level. The structure and dimensions of the equilibrated vesicle are depicted in Figure 1.

## II. Methodology

### a. Composition of the Vesicle

The head groups of the lipid molecules play an essential role in the biological functions of the SV such as the recognition and binding of synaptic proteins.[46–49] Lipidomics studies have found (1) the molar ratio of cholesterol to phospholipid is about 1:2;[30,50] (2) the three most predominant types of phospholipid molecules are phosphatidylcholine (PC), phosphatidylserine (PS), and phosphatidylethanolamine (PE);[51,52] (3) the negatively charged PS lipids mainly reside on the outer leaflet;[30] (4) most PC lipids are reported as 16:0/18:1 (1-palmitoyl-2-oleoyl-glycero-3-phosphocholine, POPC)[30]; and (5) tails of PE and PS lipids are relatively long and highly unsaturated (e.g., 20:4/20:4)[30,53]. Ideally, such lipid molecules should be employed in constructing the vesicles. However, short-term simulation indicates that such long-tailed lipids result in a vesicle exceeding the maximal size allowed by Anton2[54], the supercomputer that is necessary to achieve the necessary amount of sampling. This point will be further addressed in the next section. As a compromise, lipids of the same head groups with shorter, unsaturated tails, i.e., POPE (16:0-18:1) and POPS (16:0-18:1), are employed along with POPC (16:0-18:1) to construct the model vesicle. The outer leaflet of the model SV is chosen to be a 3:4:1:4 (molar ratio) mixture of POPS, POPE, POPC, and cholesterol, whereas the inner surface of the model SV is composed of a 1:7:4 (molar ratio) mixture of POPE, POPC and cholesterol. Both leaflets combined, the total ratio of POPS, POPE, POPC and cholesterol is 4:6:6:8, which has been shown to closely resemble the composition of lipids of a physiological SV.[30,50,51,55,56] The details of the model vesicle are summarized in Table S1. It is important to note that a flat membrane with an identical lipid composition (e.g., one leaflet has the same composition as the outer leaflet and the other leaflet has the same composition as the inner leaflet) is also simulated and serves as a control group to highlight the curvature-induced properties.

### b. Construction of the Simulation System

Although dispersing these lipids randomly in aqueous solution and waiting for them to naturally form a vesicle is possible in theory, this method suffers from several flaws, including losing control of the inner and outer-leaflet composition, allowing the formation of smaller micelles and/or hemi-fused vesicles,[57] and most importantly, computational inefficiency. Instead, the formation of the vesicle is precisely controlled by creating two layers of lipids whose head groups are constrained to the surface of two concentric spheres with the PACKMOL[58] software. The vesicle is of 14.5 nm diameter with a thickness of ~3.8 nm[29], is chosen to create a balance between the computational resources and length of the simulations while recreating the lower limit of SV size found in experiments[30] and previous *in-silico* studies[24]. As such, the head groups of the inner leaflet lipids are constrained to the surface of a sphere of radius 3.45 nm, and methyl carbons (acyl tails) to a sphere of 5.2 nm while the head groups of the outer leaflet are constrained to the surface of a sphere of radius 7.25 nm and the methyl carbons (acyl tails) to a sphere of 5.5 nm (Figure S1).

On a flat, single component membrane, one phospholipid molecule on average occupies a lateral area (i.e., area per lipid, APL) of approximately 60 Å$^2$,[29] and the APL increases/decreases by approximately 25-35% as a result of a similar convex/concave curvature as the ones found in the model SV.[37,57,59] Therefore, an APL of 80 Å$^2$ (~33% larger than the APL of a flat membrane) and 40 Å$^2$ (~33% smaller than the APL of a flat membrane) are employed to approximate the surface area of the outer (~ 660 nm$^2$, 825 lipids) and inner (~ 141 nm$^2$, 352 lipids) leaflet for packing purposes, respectively. The number of lipid molecules is then slightly reduced to 785 and 340 to accommodate specific ratios of lipids in each leaflet (sec. **II.a**). Cholesterol (number indicated in Table S1) are then inserted into the bilayer of the model SV.

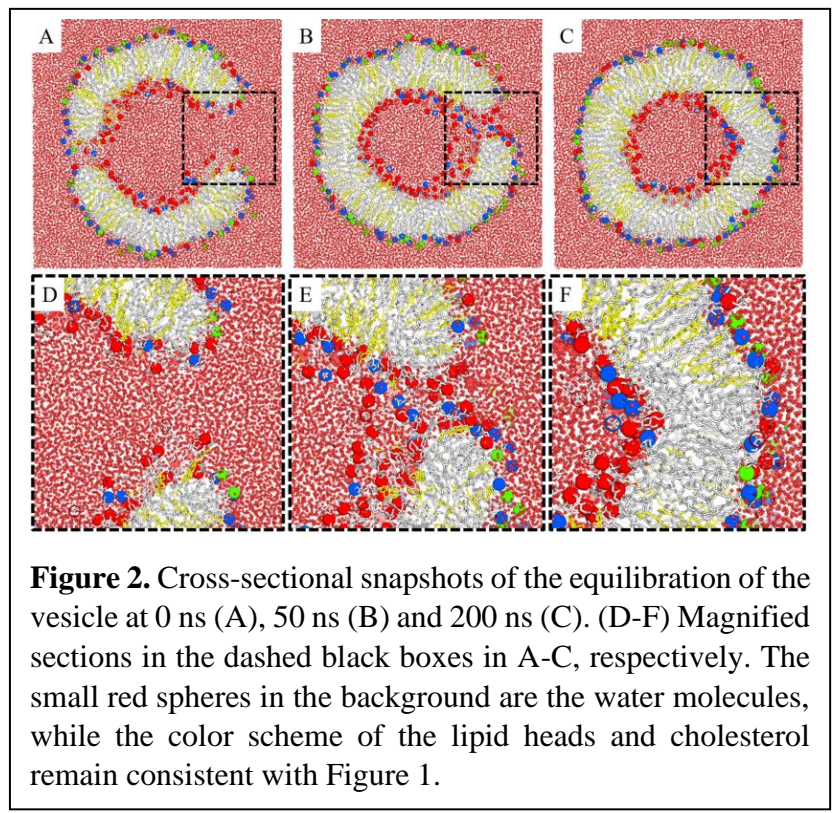

**Figure 2.** Cross-sectional snapshots of the equilibration of the vesicle at 0 ns (A), 50 ns (B) and 200 ns (C). (D-F) Magnified sections in the dashed black boxes in A-C, respectively. The small red spheres in the background are the water molecules, while the color scheme of the lipid heads and cholesterol remain consistent with Figure 1.

After constructing the vesicle, 24,000 water molecules are added to fill the cavity inside, while another 163,295 water molecules are added to the outside bound by a cubic box with a box length of 19.5 nm. It is important to note that the initial density of the water inside the vesicle is intentionally made nearly 4 times larger than outside, which is close to the density of water at 310.15 K. The purpose of the excessive packing of water inside the vesicle is to force the formation of pores,[24,60] which is used to facilitate possible shuffling of lipid molecules between both leaflets and accelerate the equilibration of the vesicle (Figure 2).[24] This procedure also prevents large undulations and distortions of the vesicle surface as well as significant variation in the structural and dynamic properties of the vesicle (such as forming prolate a spheroid shapes).[61] The final density of water within the vesicle, sphericity of the vesicle, and pore area following equilibration can be seen in Figure S2. 966 K$^+$ and 672 Cl$^-$ are introduced for physiological relevance (0.15 M) and the excess K$^+$ is used to balance the charge of the system (e.g., compensating the negative charges introduced by PS). K$^+$, instead of Na$^+$, is employed as the cation – although both cations exist in physiological presynaptic terminals, K$^+$ is the most abundant cation within the nerve cell intracellular fluid while Na$^+$ is the most abundant cation in the extracellular fluid.[62–64] Ideally, NaCl should be mostly inside of the vesicle and KCl should be mostly outside of the vesicle since SVs are recycled within the neuron.[7] However, the aforementioned procedure of constructing the model SV requires solvent exchange between the inside and outside of the vesicle, thus only KCl is added as an approximation. The system consists of a total of ~750,000 atoms in a cubic box of ~19.5 nm in each dimension. Periodic boundary conditions are applied in all directions.

### c. Molecular Dynamics simulations

The CHARMM36m[65] and TIP3P[66] forcefields are employed to model lipids and water molecules, respectively. The initial structure (sec. **II.b**) is equilibrated with Gromacs-2020.4[67] following regular MD convention: After the initial 5000 steps of steepest descent minimization, six stages of equilibration that systematically increase the time step and decrease the restraints on the lipid molecules is carried out. The first two steps (NVT ensemble) of equilibration are carried out starting with a 1 fs time step, and the Berendsen weak coupling method[68] is used as the thermostat to maintain a temperature of 310.15 K with a time constant of 1 ps. The third stage (NPT ensemble) is performed using a 1 fs time step and the Berendsen weak coupling method to maintain pressure isotropically at 1.0 bar with a compressibility constant of 4.5 x $10^{-5}$ bar$^{-1}$. For the remaining three steps, the time step is increased to 2 fs and the restraints continually decrease to zero. A 400-ns AA MD simulation is then carried out to further equilibrate the system with an NPT ensemble maintained at 310.15 K and 1.0 bar using the Nose-Hoover thermostat[69] and the Parrinello-Rahman barostat[70], respectively. The cut-off distance for the short-range nonbonded interactions is 1.2 nm, with electrostatic interactions calculated using the particle-mesh Ewald method[71]. Constraints are imposed on all the simulations using the LINCS algorithm[72]. It is important to realize the burdens of the AA MD simulation: with 2 fs as the integration step, the estimated speed with eight Nvidia 2080Ti GPUs is only 20 ns/day.[73]

To achieve a meaningful length of simulation that make the result statistically significant, a specialized supercomputer is necessary. Anton2 is the second-generation supercomputer built for biomolecular simulations that achieves significant gains in performance compared to consumer-grade hardware.[54] The caveat of Anton2 is that it only allows for a system of less than ~750,000 atoms in order to take full advantage of its computation efficiency. The final frame of the system following the 400-ns equilibration in Gromacs is ported to Anton2 for long production runs. On Anton2, the equations of motion are integrated using the Verlet algorithm with a 2.5 fs time step. The temperature was maintained at 310.15 K using the Nose-Hoover thermostat and the pressure at 1.0 atm using the Martyna−Tobias−Klein (MTK) barostat.[74] The temperature and pressure coupling are applied every 480 and every 24 simulation steps, respectively. The Lennard-Jones interactions were truncated at 9.0 Å and long-range electrostatics were computed by the k-space Gaussian split Ewald method[75] on a 64 × 64 × 64 point grid. These setting are chosen following the common protocol of Anton2. The total length of the MD simulation exceeds 10 $\mu$s.

### III. Results

Table 1 summarizes the properties of the highly curved model SV and the flat membrane of the same composition (sec. **II.a**).

**Table 1.** Radius (r), surface area (A), membrane thickness (d), area per lipid (APL), diffusion coefficient (D) and order parameter ($S_{CD}$) for the vesicle membrane and the flat membrane.

| Curvature | Leaflet | | r | A | d | APL | D | $S_{CD}$ | |
|---|---|---|---|---|---|---|---|---|---|
| | | | (nm) | (nm$^2$) | (nm) | (nm$^2$) | (x 10$^{-8}$cm$^2$/s) | (sn-1) | (sn-2) |
| Flat | Upper | Average | ∞ | 158.5 | 2.132 | 0.4459 | 2.177 | 0.3063 | 0.2348 |
| | | POPC | | | | 0.4901 | 2.435 | 0.3076 | 0.2348 |
| | | POPE | | | | 0.4748 | 2.161 | 0.3041 | 0.2333 |
| | | POPS | | | | 0.4854 | 2.099 | 0.3090 | 0.2369 |
| | | Cholesterol | | | | 0.3749 | - | - | - |
| | Lower | Average | ∞ | 158.5 | 2.036 | 0.4782 | 3.559 | 0.2577 | 0.1944 |
| | | POPC | | | | 0.5249 | 3.873 | 0.2601 | 0.1964 |
| | | POPE | | | | 0.5016 | 3.507 | 0.2572 | 0.1941 |
| | | Cholesterol | | | | 0.3907 | - | - | - |
| Vesicle | Outer | Average | 7.8 | 768.6 | 1.818 | 0.6510 | 17.25 | 0.2076 | 0.1471 |
| | | POPC | | | | 0.7410 | 17.09 | 0.2072 | 0.1471 |
| | | POPE | | | | 0.7119 | 17.41 | 0.2068 | 0.1459 |
| | | POPS | | | | 0.7130 | 17.11 | 0.2089 | 0.1489 |
| | | Cholesterol | | | | 0.5181 | - | - | - |
| | Inner | Average | 4.2 | 220.1 | 1.739 | 0.4038 | 12.17 | 0.1146 | 0.0896 |
| | | POPC | | | | 0.4392 | 12.21 | 0.1191 | 0.0929 |
| | | POPE | | | | 0.4201 | 11.93 | 0.0844 | 0.0678 |
| | | Cholesterol | | | | 0.3400 | - | - | - |

## a. Vesicle Equilibration

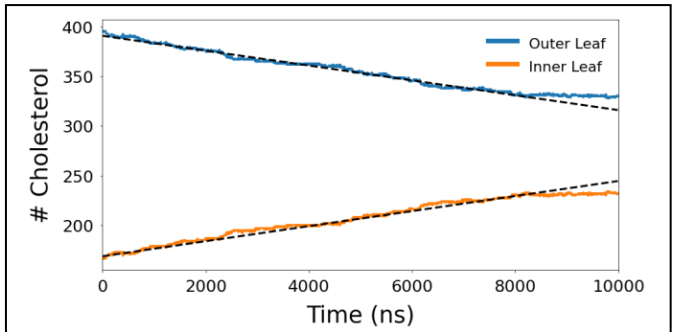

**Figure 3.** Time series of the number of cholesterols residing on the outer (blue) and inner (orange) leaflets. Dashed black lines are best-fit lines with a slope of approximately +/- 7.4 cholesterol/$\mu$s (between 0 and 8 $\mu$s).

The equilibration steps listed in sec. **II** are followed by standard MD simulation protocols that have been proven successful in modeling a flat bilayer[29,76,77], which can be validated by comparing the simulated properties (e.g., thickness and APL) with experiments. However, as stated in sec. **I**, such information on a vesicle of similar size and lipid composition is very limited from the experimental perspective and does not exist from the simulation perspective. Therefore, special cautions are granted to avoid the risk of declaring an equilibrated system too soon. Herein, the distribution of cholesterols between the outer and inner leaflets are employed as an example to demonstrate this point. As shown in Figure 3, following the initial 400 ns of equilibration (sec. **II**), the cholesterols have a net flux of transiting to the inner leaflet. It takes roughly 8 $\mu$s, with an average translocation rate of 7.4 cholesterol/$\mu$s, before the number of cholesterols reaches an equilibrium between leaflets. A flat membrane of the same composition is simulated but no net cholesterol flux is recorded in the 1 $\mu$s trajectory, which is in accordance with previous atomistic studies of flat membranes of similar conditions.[78–80] These results indicate that the initial lipid composition is stable in a flat membrane, but not stable as a result of the curvature of the vesicle. Therefore, the equilibration of the system is not achieved until 8 $\mu$s, at which point, roughly 16% of the cholesterols initially located in the outer leaflet have migrated to the inner leaflet. We reiterate that the initial lipid composition is carefully chosen from lipidomics studies (sec. **II**) – although the simulation starts from the best guess according to experiments, significant simulation time is still required to ensure proper equilibrium in the case of a highly curved membrane.

**b. Diffusion and Lipid Mixing on a Curved Surface**

Although the diffusivity of various lipid molecules has been well measured experimentally, these studies employ bilayers with significantly less curvature (e.g., giant unilamelar vesicles (GUVs) attached to a substrate). Often, a bimodal lipid diffusivity[81–85] is observed in these studies, most likely resulting from the slow diffusion of lipids near substrate and fast diffusion of lipids far from substrate.[86] Moreover, different techniques (e.g., fluorescence correlation spectroscopy[87,88], dynamic light scattering[89], or single particle tracking[87]) and substrate types[90–92], yield widely varying results for remarkably similar lipid membranes. Regarding (AA) MD simulations, they all employ a flat bilayer, which is typically symmetric, composed of a ternary mixtures of lipids (e.g., dioleyoylphosphatidylcholine (DOPC), dipalmitoylphosphatidylcholine, (DPPC), and cholesterol).[93,94] Therefore, previous studies offer little information regarding the diffusivity of phospholipids in a highly curved membrane.

Unlike in a flat membrane (i.e., a vesicle of infinite radius), all atoms of each lipid in a vesicle *on average* move at the same *angular* speed but differ in their *linear* velocity – those atoms that are further away from the center of the vesicle move at a larger *linear* velocity than those atoms that are closer to the center. This effect can be seen in Figure S3.A in the Supporting Information. The lateral diffusion coefficient is computed from the angular mean-square displacements[95] (AMSD) scaled by their corresponding radius, which is determined with respect to the inner and outer spherical surface made by the center of mass (COM) of the phospholipids. For short periods of time (less than 100 ns) the average AMSD linearly increases with the time, $t$, which is computed as

$$\langle \theta^2 \rangle \cong \frac{2D_T}{r^2} t \qquad (1)$$

where $\theta$ is the angle between the vector connecting the COM of the vesicle and the COM of a given phospholipid at $t = 0$ and the vector of the same lipid at time $t$. $D_T$ is the diffusion coefficient, $r$ is the radius of the corresponding spherical surface. It is important to note that there is no rotational restriction applied to the vesicle, thus it could tumble throughout the simulation.

This issue is remedied by calculating the optimal rotation matrix that minimizes the RMSD of the vesicle between consecutive frames, effectively removing the overall rotational of the vesicle (see Figure S4 in the Supporting Information). The diffusion coefficient of the lipids of the SV is shown in Table 1. It is also important to note that the diffusion coefficient of Brownian particles on a spherical surface should converge to:[95]

$$\langle \theta^2 \rangle = \frac{\pi^2 - 4}{2} \quad (2)$$

As shown, the lipid molecules of the vesicle indeed demonstrate such random motion (Figure S5).

As a comparison, the diffusion coefficient of lipids is also computed with a flat bilayer of the same composition, which is shown to be on par with experiments.[76,92,96] Table 1 shows that the lipids diffuse at higher rates in a highly curved environment, including both the inner (concave) and outer (convex) leaflet. This difference is also reported by Przybylo et al.,[97] where the diffusion rates measured with giant unilamelar vesicles (i.e., curved membranes) were found to be more than two times faster than those with supported lipid bilayers (i.e., flat membranes). Furthermore, in even smaller phospholipid-containing organelles such as fast-tumbling isotropic bicelles, the diffusion rates are an additional two times larger than those measured in the giant unilamelar vesicles (i.e., more than 4 times larger than in supported bilayers).[98] Therefore, the 5~7 times faster diffusion rate of phospholipids measured in a highly curved vesicle than in a flat membrane calibrates well with experiments. Previous AA MD simulations of a flat membrane have reported that lipid mixing occurs on a time scale of microseconds,[32] which was determined with a 100 nm$^2$ patch of phase separated lipids (a ternary mixture of DOPC/DPPC/cholesterol).[99] Given the lipids on a flat membrane have a diffusion coefficient larger than 2 x 10$^{-8}$ cm$^2$/s,[96,97,100] lipids in a 10 $\mu$s simulation would cover an area of over 200 nm$^2$. As shown in Table 1, lipids in the model SV diffuse 5~7 times faster, thus with a 10 $\mu s$ simulation, the lipids could cover an area over 1000 nm$^2$. This evidence supports that the simulation is sufficient to observe lipid mixing of the vesicle (surface area of 767 and 220 nm$^2$ for the outer and inner leaflets, respectively).

c. Area Per Phospholipid (APL) and Solvent Accessible Surface Area (SASA)

To characterize the surface of the vesicle, a Voronoi diagram (Figure 1.B) is sketched according to the phospholipid's phosphorus atom or cholesterol's hydroxyl oxygen. The Voronoi vertices and edges define a Voronoi cell, which is employed to estimate the APL of phospholipids and cholesterol. For a perfect sphere, the area of each Voronoi cell may be exactly calculated. However, along the simulation, the vesicle deforms from a perfect sphere, thus Shepard's interpolation is applied to the vertices to better resemble the surface texture of the vesicle. After a triangular mesh of the surface, the area of each Voronoi cell is then estimated as the sum of the areas of the $k$ triangles composing it (Figure S6). The results of the APL of the inner and outer leaflets of the vesicle are shown in Table 1. The validity of the method can be attested with a flat membrane, whose computed APL (0.47 nm$^2$) agrees well with the experiments of comparable amounts of cholesterol (0.45 nm$^2$).[101,102] Compared to a flat bilayer, the convex curvature (i.e., outer leaflet) increases the APL by ~30%, while the concave curvature (i.e., inner surface) only slightly decreases the APL by 9%. In a recent AA MD study, Yesylevskyy et al. developed a protocol that precisely bends a flat bicelle to arbitrary curvatures to mimic a highly curved membrane. Our research indicates that Yesylevskyy et al. represent a similar variation (e.g., ~30% increase) of APL for convex side of the bent bicelle, but overestimate the impact of the concave side (e.g., ~30% decrease).[37,103]

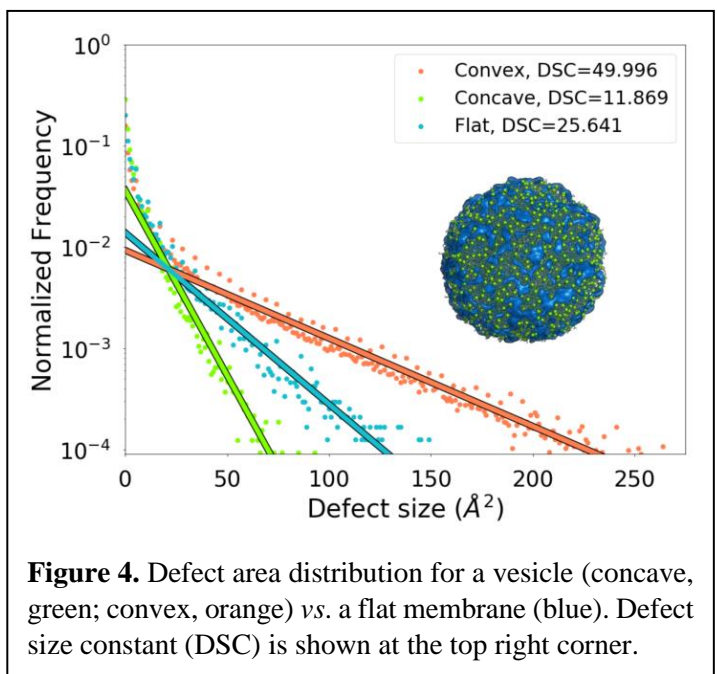

**Figure 4.** Defect area distribution for a vesicle (concave, green; convex, orange) *vs.* a flat membrane (blue). Defect size constant (DSC) is shown at the top right corner.

Lipid membrane packing defects (PDs), which are essential for the regulation of peripheral proteins' binding to the membrane, have been well-studied with near-flat lipid bilayers in MD simulations.[104] Our simulation of the vesicle provides the first opportunity to probe into the interrelationship between the curvature and PDs. The commonly adapted accessible surface area (ASA), defined as the accessible areas in the head group region that are large enough for the lipid tails to contact hydrophobic protein side chains, is employed to characterize the PDs.[41,105] More details of computing ASA can be found in the Supporting Information. The impact of the curvature on PDs is demonstrated in Figure 4. The convex curvature of the outer surface induces large PDs while the concave curvature of the inner leaflet suppresses large PDs. The flat membrane hosts a number of large PDs that are between the concave and convex membrane. Additionally, the results show that a single parameter – the defect size constant (DSC), which is the inverse of the exponential decay constant, is capable of quantitatively capturing the PDs of the membrane.[104,106] Although no PDs of a vesicle have been investigated using AA MD, there have been studies on a bent membrane as a results of bar domain proteins[41]. The DSC reveal the same trends observed, e.g., decreased defect sizes for the concave and increased defect sizes for the convex in comparison to flat membranes. The PDs reported in Figure 4 overall appear to be larger than previous studies[106,107], which may be attributed to the large curvature of the vesicle.

### d. Curvature and Lipid Packing

The AA MD simulation of the vesicle provides a means to probe into the tail structure of a highly curved membrane. The lipid order parameter ($S_{CD}$), defined by the following equation,

$$S_{CD}(n) = \frac{|\langle 3\cos^2(\theta(n))-1\rangle|}{2} \qquad (3)$$

describes the average orientation of the carbon-hydrogen bonds in phospholipid acyl tails. $\theta(n)$ is the average of the angles between the normal vector of the membrane, which varies from lipid to lipid, and the vector (e.g., 2 for methylenes, 3 for methyls) connecting the $n^{th}$ carbon in the lipid tail and the hydrogens that it is bound to. The result of the order parameters is shown in Figure S7 and more discussion on the lipid packing can be found in sec. **IV**.

### e. Membrane Phase Separation in a Curved Membrane

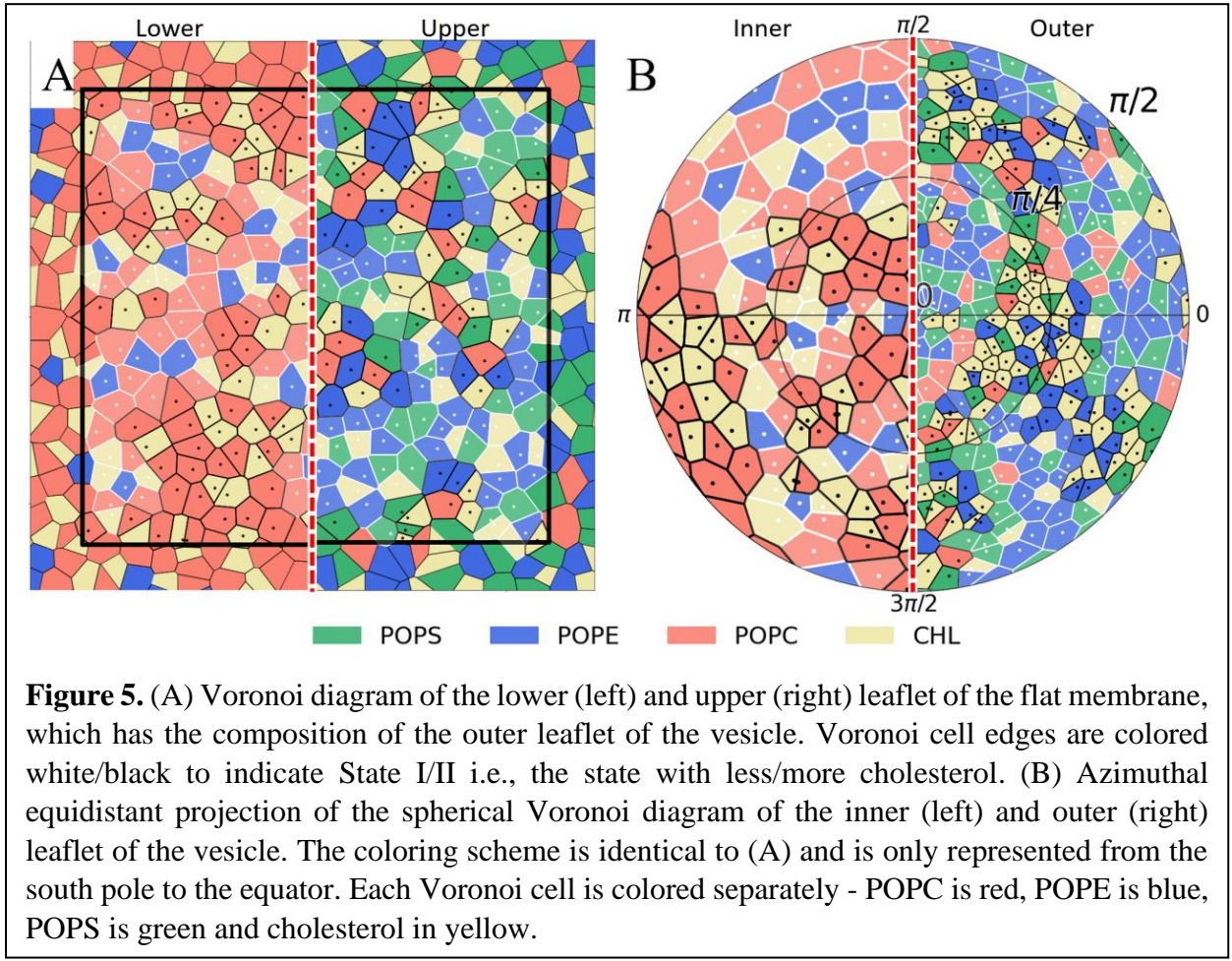

**Figure 5.** (A) Voronoi diagram of the lower (left) and upper (right) leaflet of the flat membrane, which has the composition of the outer leaflet of the vesicle. Voronoi cell edges are colored white/black to indicate State I/II i.e., the state with less/more cholesterol. (B) Azimuthal equidistant projection of the spherical Voronoi diagram of the inner (left) and outer (right) leaflet of the vesicle. The coloring scheme is identical to (A) and is only represented from the south pole to the equator. Each Voronoi cell is colored separately - POPC is red, POPE is blue, POPS is green and cholesterol in yellow.

It has been shown that in a nearly flat bilayer[99], a three-component lipid mixture containing cholesterol at ambient temperature and pressure could display various phases (e.g., gel, liquid order, and liquid disorder phases) and/or phase coexistence.[108] There have been extensive studies[32,109–114] elaborating on the relation between the local lipid composition (LLC) and their phase behavior. Moreover, there is a growing number of experiments indicating curvature and LLC are intertwined – curvature imposes geometric constraints, which in turn influences the packing of lipids.[115–117] Due to the aforementioned challenges involved in simulating a highly curved membrane, AA MD simulations have only been able to study phase behavior of a flat membrane. According to the pioneering work of Sodt et al.[94,118], liquid-ordered/liquid-disordered ($L_o/L_d$) coexistence is composed of regions of saturated hydrocarbon chains packed with local hexagonal order, separated from regions enriched in cholesterol and unsaturated hydrocarbon chains. The balance of cholesterol-rich to local hexagonal order is proposed to control the partitioning of membrane components into the liquid order regions. The current simulation of a full-scale vesicle provides an opportunity to investigate the impact of curvature on the phase behavior of a lipid bilayer.

A Hidden Markov state model[119] (HMM) is employed to characterize the phase of the vesicle (i.e., the hidden states). Inspired by previous studies[94], the LLC, defined as the five closest neighboring molecules of each lipid molecule, is chosen to be the observables of the HMM. Each phospholipid and cholesterol molecule are assumed to be in one of the two hidden states and can possibly emit any one of 84 signals if in the outer leaflet (4-component, see Figure S8) or 28 signals if in the inner leaflet (3-component) with distinct emission probabilities. The two hidden states could convert to one another or maintain themselves with distinct transition probabilities. The training data for the HMM is the time series of the LLCs for each lipid and cholesterol over a total of approximately 8000 frames obtained from the 2.0 $\mu s$ simulation after the equilibration. The Baum-Welch algorithm[120] is employed to solve for the probabilities that maximize the likelihood of observing the time series of the LLC for all the lipid molecules. HMMs with hundreds of varying initial conditions (e.g., different HMM parameters such as guesses of initial hidden states, transition and emission probabilities) have been attempted (Figure S9) and the converged HMM parameters are selected with two criteria: 1) overall largest probability of emitting the time series of LLCs and 2) most initial conditions converge to.

**Table 2.** Composition of the two states in the vesicle and flat membrane determined by HMM

|  | Flat (Upper) | | Flat (Lower) | | Vesicle (Outer) | | Vesicle (Inner) | |
| --- | --- | --- | --- | --- | --- | --- | --- | --- |
|  | *State I* | *State II* | *State I* | *State II* | *State I* | *State II* | *State I* | *State II* |
| %POPC | 20.13 | 0.931 | 53.45 | 60.65 | 10.26 | 9.296 | 51.08 | 53.79 |
| %POPE | 30.01 | 36.94 | 19.19 | 0.897 | 43.46 | 24.39 | 18.19 | 0.476 |
| %POPS | 19.41 | 30.53 | -- | -- | 33.37 | 17.52 | -- | -- |
| %Cholesterol | 30.45 | 31.59 | 27.34 | 38.45 | 12.89 | 48.79 | 30.72 | 45.73 |

Cholesterol is well-known for decreasing the fluidity of a membrane and its composition could serve as an indicator of possible phase separation.[32,102,118,121–126] The HMM determines that the curvature of the membrane could introduce two different hidden states associated with vastly different cholesterol compositions (Table 2). In the convex surface (i.e., outer leaflet), there are almost four times as many cholesterols found in one hidden state (48%) as in the other (13%). Therefore, it is *hypothesized* that the hidden states with high cholesterol composition to be the $L_o$ state and the hidden states with low cholesterol composition to be the $L_d$ state. Representative snapshots of the coexistence of these states are depicted in Figure 5. Unlike a flat membrane, it is difficult to visualize the entire vesicle surface, as all equal-area maps have distortions. The azimuthal equidistant projection is employed, which accurately represents area adjacent to a chosen point (e.g., the north or the south pole of the sphere). As the figure shows, on the convex surface, molecules of the same state tend to cluster instead of distributing randomly. Regarding the concave surface (i.e., inner leaflet), the difference in cholesterol composition between the two states is much smaller, e.g., 46% *vs*. 31%. Therefore, it is more difficult to *hypothesize* that the concave curvature introduces phase separation and further analysis will be presented in the next section. In comparison, a flat membrane of the same composition as the vesicle is simulated but the same HMM analysis renders very different results (Table 2 and Figure 5) – the two states have similar cholesterol composition, instead, the largest difference is found in POPC (upper) and POPE (lower). It is important to note that although POPC, POPE and POPS differ by head group composition, these lipids share the same tail, thus similar phase behavior are expected (i.e., comparable melting temperatures).[93] Therefore, in spite of the same composition, the level of cholesterol enrichment is not observed in a flat bilayer as in a highly curved bilayer (especially in the convex curvature). [127–129]

## IV. Discussion

**Table 3.** Simulated vesicle and flat membrane average phase separated properties

|  | APL | D | $S_{CD}$ | |
|---|---|---|---|---|
|  | (nm$^2$) | (x 10$^{-8}$cm$^2$/s) | (sn1) | (sn2) |
| Flat (Outer) | | | | |
| **State I** | 0.4548 | --* | 0.3064 | 0.2349 |
| **State II** | 0.4413 | --* | 0.3104 | 0.2393 |
| Flat (Inner) | | | | |
| **State I** | 0.4891 | --* | 0.2604 | 0.1956 |
| **State II** | 0.4677 | --* | 0.2650 | 0.1986 |
| Vesicle (Outer) | | | | |
| **State I** | 0.7736 | 19.69 | 0.1940 | 0.1364 |
| **State II** | 0.5564 | 13.84 | 0.2377 | 0.1708 |
| Vesicle (Inner) | | | | |
| **State I** | 0.4392 | 10.85 | 0.0933 | 0.0678 |
| **State II** | 0.4201 | 10.01 | 0.1288 | 0.0929 |

* Insufficient sampling of continuous lipid sequences in State I or State II to provide an adequate estimate of diffusivity

The AA MD simulation of the full-scale vesicle has revealed the impact of curvature on the sorting of lipids that could potentially lead to phase coexistence. Previous experiments and simulations[130–132] have investigated curvature-induced phase separation with simple ternary mixtures of lipids (e.g., phosphatidylcholine-based phospholipids, cholesterol, and sphingomyelin) at a temperature lower than the lipid phase-transition temperature ($T_m$). The present work, to our knowledge, is the first example of curvature inducing phase separation of a quaternary mixture of lipids at $T > T_m$ with AA MD. It is important to acknowledge that special cautions are necessary before blindly accepting the HMM-assigned states, as according to the HMM, the lipids must be categorized into distinct, discontinuous states. Thus, to ensure that those HMM-assigned states are meaningful, and the curvature-induced phase coexistence is real, properties that are related to the phase of lipids

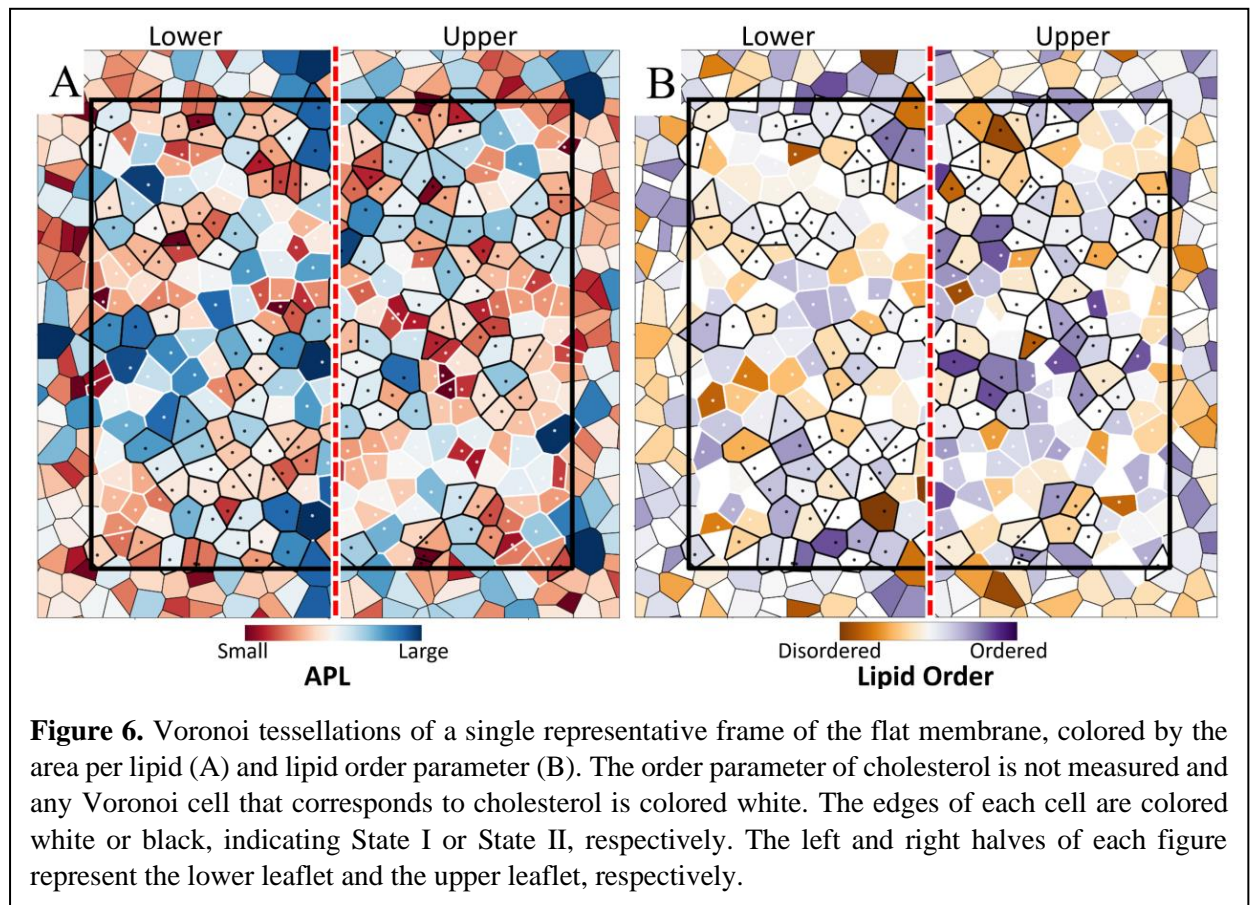

**Figure 6.** Voronoi tessellations of a single representative frame of the flat membrane, colored by the area per lipid (A) and lipid order parameter (B). The order parameter of cholesterol is not measured and any Voronoi cell that corresponds to cholesterol is colored white. The edges of each cell are colored white or black, indicating State I or State II, respectively. The left and right halves of each figure represent the lower leaflet and the upper leaflet, respectively.

are discussed.

APL and lipid order are deemed crucial biophysical indictors of a membranes phase.[32,133-135] In the case of the flat membrane (Table 3), the average APL and lipid order of molecules belonging to two different states are nearly indistinguishable. This phenomenon can also be seen in Figure 6.A and 6.B, where molecules of different APL and order parameters are randomly distributed into two different phases. To further investigate the possibility of phase separation, the lifetimes ($t_R$) of lipids remaining in the same state are analyzed. Lipid molecules are categorized as core (Figure 7.A) and non-core (Figure 7.B) lipids depending on their neighboring lipids – core lipids are surrounded by lipids of the same state, while non-core lipids should have some neighboring lipids of a different state. The density based spatial clustering of applications with noise (DBSCAN)[136] is employed for the categorization of core vs. non-core lipids. In a well-equilibrated lipid phase coexistence, lipids switch states (i.e., phases) on the boundaries between the two states. Therefore, if there were phase coexistence and the two phases occupy roughly equal space, core lipids belonging to the more disordered phase are expected to have smaller $t_R$ compared to core lipids belonging to the more ordered phase, as the former is more likely to travel to the boundaries due to high diffusivity. As Figure 8.A and 8.B shows, the distribution of $t_R$ of both states of the flat membrane are very similar to one another, indicating that lipids belonging to both states are diffusing at a very similar rate. Considering all the aforementioned evidence (including the lipid composition in Sec. **III.e**), it appears that there is no phase separation in either upper or lower leaflets of the flat membrane.

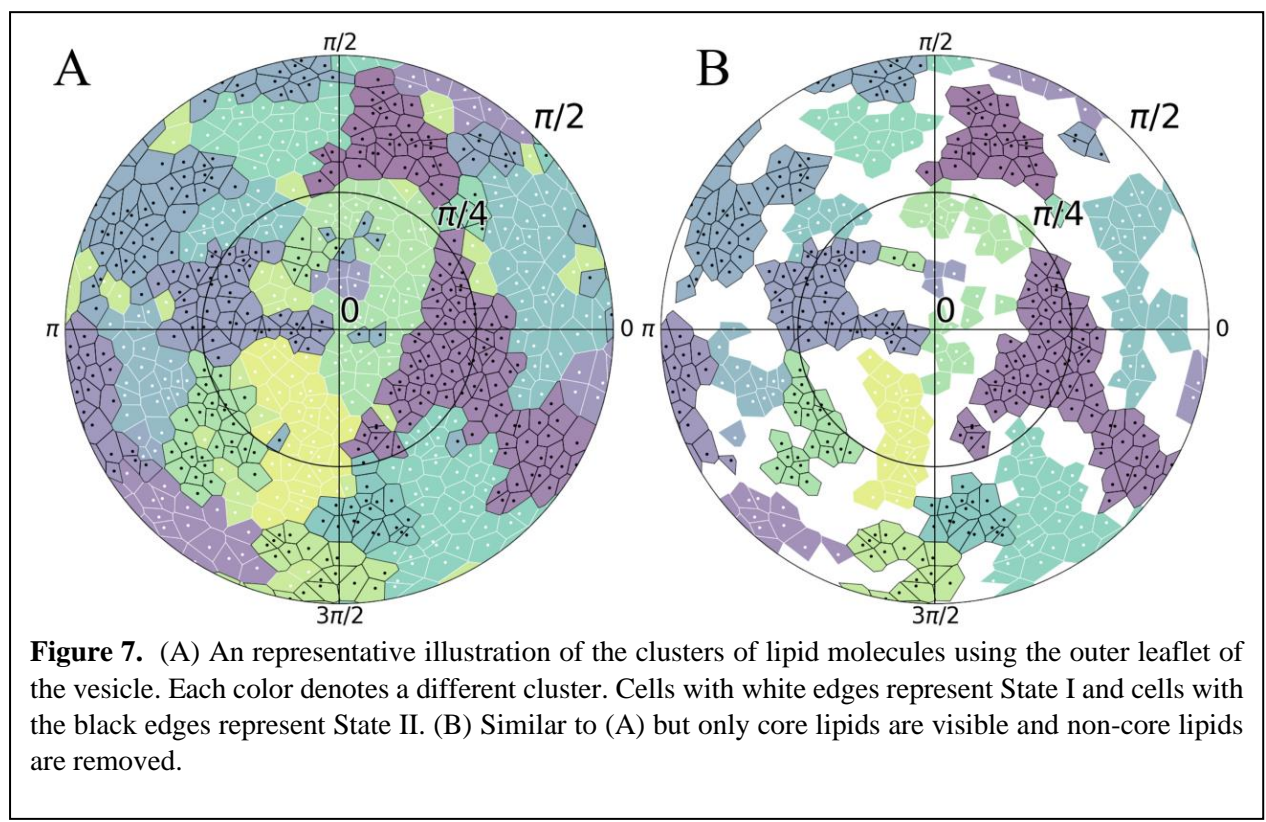

**Figure 7.** (A) An representative illustration of the clusters of lipid molecules using the outer leaflet of the vesicle. Each color denotes a different cluster. Cells with white edges represent State I and cells with the black edges represent State II. (B) Similar to (A) but only core lipids are visible and non-core lipids are removed.

Compared to a flat bilayer, the lipid molecules of two states found in a highly curved bilayer (i.e., the vesicle membrane) have vastly different properties. Regarding the outer leaflet (convex curvature), the composition of the two states is summarized in Table 2 and their corresponding properties are summarized in Table 3. The average APL of molecules of State II are roughly 30% smaller than those of State I, indicating that lipid heads of State II are more densely packed than molecules of State I (i.e., less PDs). Characterized by the order parameter ($S_{CD}$), there are also significant differences in lipid tail packing – molecules in State II are more ordered (~ 20%) than those in State I. More importantly, compared to the case of a flat membrane (Figure 6), Figure 9 shows that molecules with distinct APL or order parameters aggregate and are categorized into distinct states by the HMM, e.g., most of the molecules with small APL (colored in red) aggregate and make a cluster of State II. Similar $t_R$ analysis is carried out for core lipids as well, and the result (Figure 8.C) shows that core lipids in State I have a significantly shorter lifetime of retaining the same state compared to core lipids in State II. In other words, the simulation suggests that State II is structurally stable while State I is highly dynamic – molecules travel to the edge more frequently and are more likely to switch phases. This result also agrees with the calculated diffusivity: lipids in State I diffuse 30% faster compared to those in State II (Table 3). Although experiments using the asymmetric phospholipid and cholesterol compositions as seen here are exceedingly rare, simple POPC and cholesterol mixtures are available. Experimentally, the ratio of the diffusion coefficient for POPC with 10% and 50% cholesterol[137] is ~0.63, in good agreement to the ratio of ~0.70 found here between State I (13% cholesterol) and State II (49% cholesterol). As State I and II are spatially separated, consisting of vastly different amounts of cholesterol, experiencing different levels of defects and packing, and are diffusing at different rates, it is concluded that the $L_o$ and $L_d$ phases coexist in the highly convex surface, i.e., the outer leaflet of the vesicle.

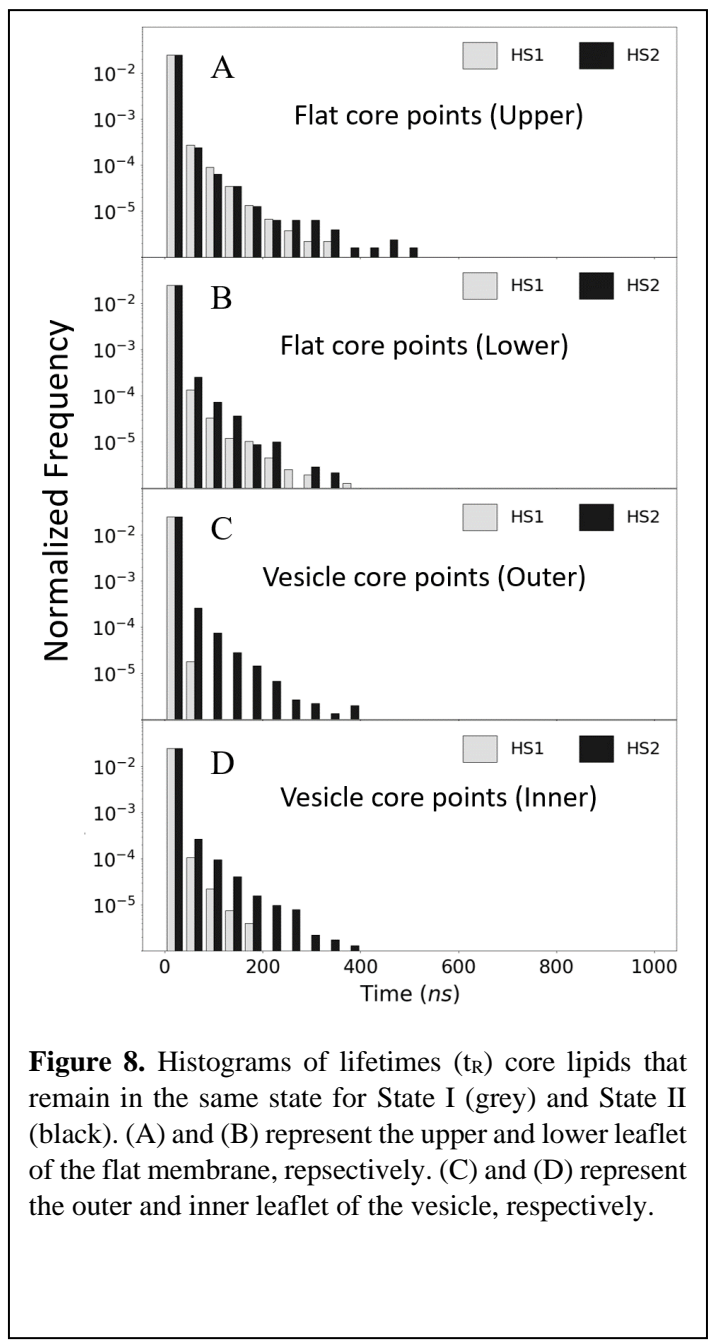

**Figure 8.** Histograms of lifetimes ($t_R$) core lipids that remain in the same state for State I (grey) and State II (black). (A) and (B) represent the upper and lower leaflet of the flat membrane, repsectively. (C) and (D) represent the outer and inner leaflet of the vesicle, respectively.

Compared to the outer leaflet (convex curvature) of the vesicle, whether there is phase coexistence in the inner leaflet (concave curvature) is much less clear. The properties of molecules belonging to the two states are summarized in Table 3. Although State II consists of roughly 50% more cholesterol (46% vs. 31%, Table 2), its average APL is almost identical to State I. Further, Figure 9.A shows that molecules of distinct APL do aggregate, but they tend to be categorized into both states. These findings suggest that there is *not* a phase coexistence in the inner leaflet. Order parameters, on the other hand, indicate that the molecules in State II are packed more orderly than those in State I (Table 3), although their spatial separation is somewhat not clear (Figure 9B). The analysis of the lifetimes ($t_R$) of lipids remaining in the same state is also employed as an attempt to settle the conflicting information from APL and Order parameters, but unfortunately the result in Figure 8.D show that the level of differences in $t_R$ between the two states lies in between the outer leaflet (i.e., phase coexistence) and the flat membrane (i.e., no phase coexistence). The diffusion coefficient calculation does not provide strong evidence in this dispute either, as molecules in State I diffuse only slightly (8%, Table 3) faster than those in State II. Weighing in all the evidence for and against phase coexistence in the inner leaflet, it is not conclusive enough that there is phase coexistence in the inner leaflet of the vesicle.

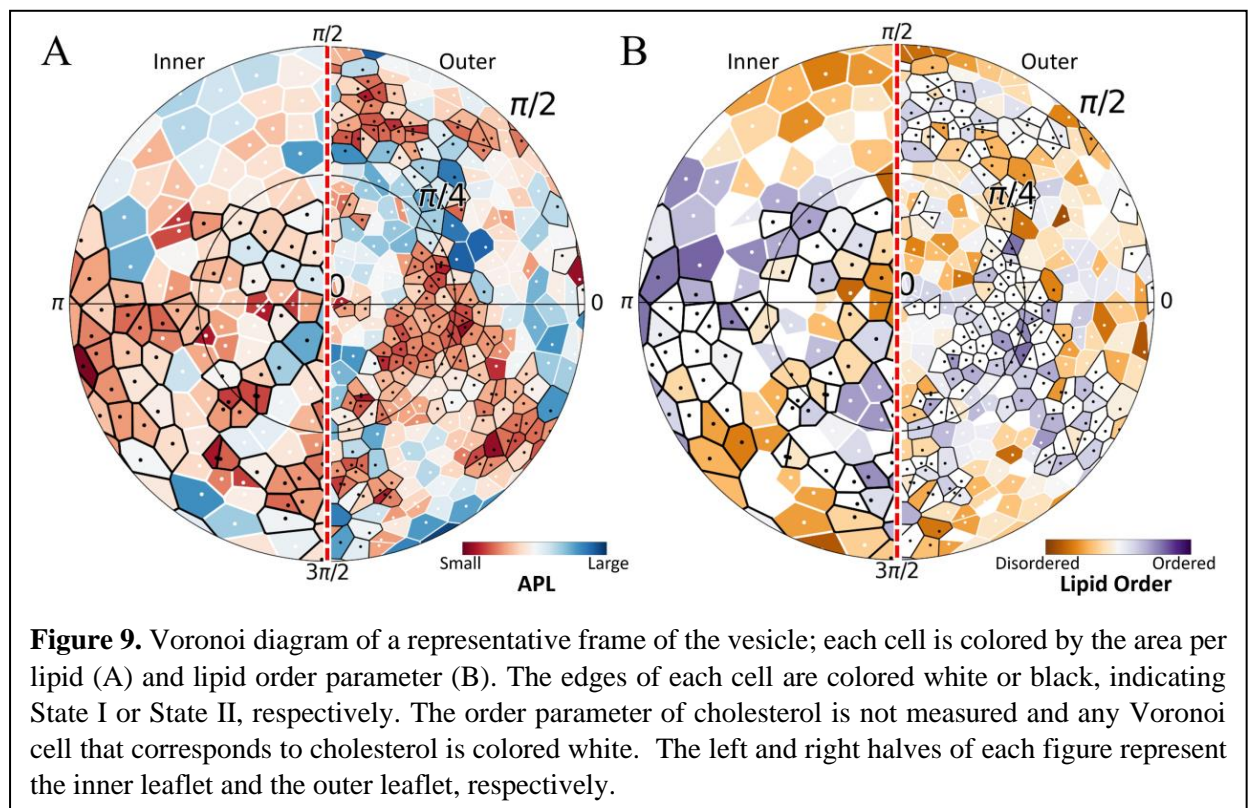

**Figure 9.** Voronoi diagram of a representative frame of the vesicle; each cell is colored by the area per lipid (A) and lipid order parameter (B). The edges of each cell are colored white or black, indicating State I or State II, respectively. The order parameter of cholesterol is not measured and any Voronoi cell that corresponds to cholesterol is colored white. The left and right halves of each figure represent the inner leaflet and the outer leaflet, respectively.

It is still of interest to investigate the structure of inner leaflet as it has demonstrated distinctively low order parameter compared to the outer leaflet as well as the flat membrane (Table 3 and Figure S7). The inner leaflet is also the thinnest among those measured in this study (Table 1). A snapshot of a representative configuration is depicted in Figure 10.A, in which the acyl tails of the lipid molecules in the inner leaflet appear to tilt and fold back on themselves. This phenomenon contributes to a small order parameter associated with the inner leaflet. To quantitatively analyze the source of such behavior, two angles are calculated for each lipid (Figure 10.B). The first angle is the tilting angle ($\theta$) of the phospholipid plane, defined as the angle between the cross product $C_{sn1} \times C_{sn2}$ and the normal of the vesicle surface, and the second angle is the splitting angle ($\phi$) of the tails, defined as the angle between $C_{sn1}$ and $C_{sn2}$. Figure 10.C suggests that the two acyl chains are split by very similar angles in both leaflets, but as shown in Figure 10.D, lipid molecules in the inner leaflet are more likely be found at a much larger tilted angles compared to the ones in the outer leaflet lipids, which most likely orient themselves parallel with the normal of the vesicle surface. It is also interesting to note that lipid molecules in the inner leaflet do not tilt in a uniform matter, but instead, randomly with respect to the normal of the vesicle surface, which results in small order parameters found in the inner leaflet. A similar acyl chain flexibility and therefore low order parameter was observed in CG MD simulations by Risselada et al.[24], who found a similar decreased order parameter, particularly in the inner vesicle leaflet.

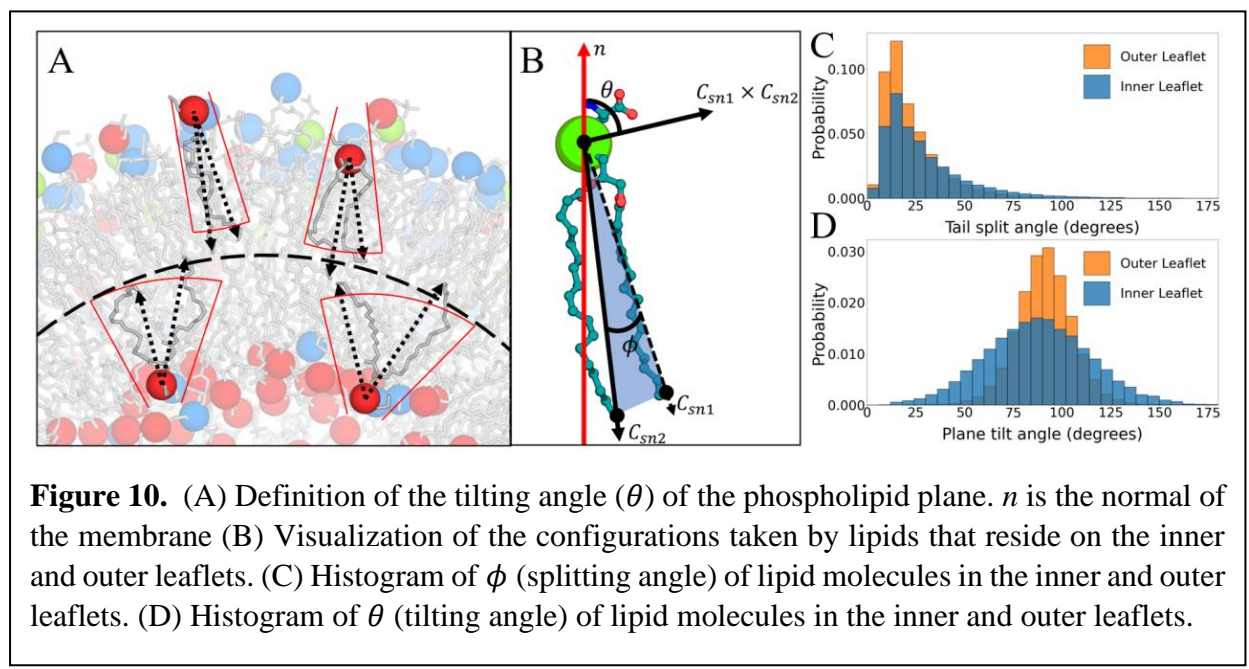

**Figure 10.** (A) Definition of the tilting angle ($\theta$) of the phospholipid plane. $n$ is the normal of the membrane (B) Visualization of the configurations taken by lipids that reside on the inner and outer leaflets. (C) Histogram of $\phi$ (splitting angle) of lipid molecules in the inner and outer leaflets. (D) Histogram of $\theta$ (tilting angle) of lipid molecules in the inner and outer leaflets.

## V. Conclusions

The (AA) MD simulation of a full-scale vesicle has revealed the interplay between the curvature and properties of lipid bilayers. Most interestingly, lipids of the same composition demonstrating no phase coexistence in a flat bilayer show very strong indictors of phase coexistence on a highly convex surface. The present work, to our best knowledge, is the very first demonstration of such phase coexistence observed in AA MD. As a result, the common biophysical properties of various lipid molecules (diffusion coefficients, surface areas per lipid, order parameter) and packing defects in a highly curved environment are characterized. Previous AA MD simulations on microscopic phase coexistence almost always employ a near or completely flat membranes, ignoring the role that various morphology might play in the phase behavior.[94,118,138,139] Further, previous (AA) MD simulations were mostly carried out at a temperature that is between various $T_m$ of the different lipid molecules. This temperature setting could give the bilayer an inherent potency for phase separation, as certain lipids tend to exist in one phase while others tend to exist in another phase.[29,137]

Live cell membrane heterogeneity most likely exists on the nanoscale under physiological conditions rather than in large-scale phases (i.e., microdomains).[140–143] Thus, the fundamental understanding of nano-scale organization of membranes is of great importance to many outstanding challenges. For example, in understanding how the curvature of synaptic vesicles affects the binding preference and therefore function of neuronal proteins such as α-synuclein, the protein solely responsible for the amyloid aggregates (Lewy bodies) in Parkinson's Disease.[144,145] In the pharmaceutical industry, highly curved membranes found in exosomes have recently been a targeted field for their potential as next-generation drug delivery vehicles.[146-150] The on/off loading of their original cargo centers around the dynamics of their highly curved membranes and how the curvature impacts the partitioning of small bioactive molecules.[146,148,150-152] Furthermore, the current study may be used as a benchmark for developing bottom-up coarse-grained models as well as designing curvature-producing biasing forces in modeling realistic gaussian curvatures.

**Supporting Information**
Additional information about the vesicle system and methods including the equilibration procedure and the HMM can be found at (_DOC)

**Acknowledgement**
This research is supported by Eli Lilly and Company. The Anton2 computer time was provided by the Pittsburgh Supercomputing Center (PSC) through grant MCB200078P from the National Research Council (NRC) at the National Academies of Science. The Anton2 machine at PSC was generously made available by D. E. Shaw Research. The authors also appreciate the Information and Technology Services (ITS) from the University of Hawai'i, Manoa, and XSEDE for the computational resources.